\documentclass[aps,prd,english,10pt,preprintnumbers,floatfix,showpacs,superscriptaddress,nofootinbib,showkeys]{revtex4-1}
\pdfoutput=1
\usepackage{lmodern}
\usepackage[T1]{fontenc}
\usepackage{textcomp}
\usepackage[scr=boondoxo,frak=boondox,bb=boondox]{mathalfa}
\usepackage[pdftex, includehead, includefoot, centering]{geometry}
\usepackage[utf8]{inputenc}
\usepackage{amsmath, amssymb, amsfonts, amsthm}
\usepackage{graphicx}
\usepackage{slashed}
\usepackage[table]{xcolor}
\usepackage[english]{babel}
\usepackage{epsfig}
\usepackage[colorlinks=true, urlcolor=blue, anchorcolor=blue, citecolor=blue, filecolor=blue, linkcolor=blue, menucolor=blue, linktocpage=true, pdfa=true]{hyperref}
\usepackage{color}

\graphicspath{{figs/}}

\catcode`\@=11
\def\lsim{\mathrel{\mathpalette\@versim<}}
\def\gsim{\mathrel{\mathpalette\@versim>}}
\def\@versim#1#2{\vcenter{\offinterlineskip
\ialign{$\m@th#1\hfil##\hfil$\crcr#2\crcr\sim\crcr }}}
\catcode`\@=12

\newcommand{\AddrUNAM}{Instituto de Física, Universidad Nacional Autónoma de México, A.P. 20-364, Ciudad de México 01000, México.}
\newcommand{\AddrUCN}{Departamento de Física, Universidad Católica del Norte, Avenida Angamos 0610, Casilla 1280, Antofagasta, Chile.}
\newcommand{\AddrTUM}{Physik-Department T30d, Technische Universität München. James-Franck-Strasse, 85748 Garching, Germany.}
\newcommand{\AddrMCTP}{ {\it Mesoamerican Centre for Theoretical Physics -- UNACH, ctra. Zapata km 4, 29040, Tuxtla Guti\'errez, M\'exico}}

\begin{document}
\title{Fermion Dark Matter and Radiative Neutrino Masses from Spontaneous Lepton Number Breaking}
\author{Cesar Bonilla}\email{cesar.bonilla@tum.de}\affiliation{\AddrTUM}\affiliation{\AddrUCN}
\author{Leon M.G. de la Vega}\email{leonm@estudiantes.fisica.unam.mx}\affiliation{\AddrUNAM}
\author{J. M. Lamprea}\email{jorge.mario@unach.mx}\affiliation{\AddrUNAM} \affiliation{\AddrMCTP}
\author{Roberto A. Lineros}\email{roberto.lineros@ucn.cl}\affiliation{\AddrUCN}
\author{Eduardo Peinado} \email{epeinado@fisica.unam.mx}\affiliation{\AddrUNAM}


\begin{abstract}
In this paper, we study the viability of having a fermion Dark Matter particle below the TeV mass scale in connection to the neutrino mass generation mechanism. The simplest realization is achieved within the scotogenic model where neutrino masses are generated at the 1-loop level. Hence, we consider the case where the dark matter particle is the lightest $\mathbb{Z}_2$-odd Majorana fermion running in the neutrino mass loop. We assume that lepton number is broken dynamically due to a lepton number carrier scalar singlet which acquires a non-zero vacuum expectation value. In the present scenario the   Dark Matter particles can annihilate via $t$- and $s$-channels. The latter arises from the mixing between the new scalar singlet and the Higgs doublet. We identify three different Dark Matter mass regions below 1 TeV that can account for the right amount of dark matter abundance  in agreement with current experimental constraints. We compute the Dark Matter-nucleon spin-independent scattering cross-section and find that the model predicts spin-independent cross-sections ``naturally''  dwelling below the current limit on direct detection searches of Dark Matter particles reported by XENON1T.
\end{abstract}

\maketitle

\section{Introduction}
\label{sec:Intro}

The observed fundamental particles as well as their interactions via the strong and electroweak forces are well described under the Standard Model (SM) picture. However, the SM predicts massless neutrinos contradicting neutrino oscillation experiments which indicate that at most one active neutrino can be massless~\cite{Whitehead:2016xud, Decowski:2016axc, Abe:2017uxa, deSalas:2017kay, Capozzi:2016rtj,Esteban:2018azc}.
In addition, so far there is no experimental evidence on the exact mechanism chosen by
nature to generate  neutrino masses. In this regard, the most popular idea to circumvent this
mismatch between the SM and neutrino oscillation data is to assume that neutrinos are
Majorana particles and invoke the so-called $seesaw$
mechanism~\cite{Minkowski:1977sc, Yanagida:1979as, Mohapatra:1979ia, Schechter:1980gr, Schechter:1981cv, Foot:1988aq}.
Furthermore, the SM does not provide a candidate to account for the dark matter (DM) relic abundance
in the Universe. The dark matter constitutes about 80\% of the matter content
of the Universe and its presence is strongly supported by observational evidence at multiple scales,
through gravitational effects, its role in structure formation and influence in the features of the
Cosmic Microwave Background (CMB). By looking at the CMB and other observables, the Planck collaboration has
put the following limit on the dark matter relic abundance~\cite{Aghanim:2018eyx},
\begin{equation}
\label{eq:omega}
\Omega_{c} h^2 = 0.1200 \pm 0.0012 \ \ \text{at} \ \ 68\% \, \text{C.L}.
\end{equation}

Theoretically, it is very tempting to think that the DM sector and neutrino mass generation mechanism are
linked. This connection appears naturally when the neutrino masses are generated at the loop
level~\cite{Ma:2006km}.  In such scenarios, the smallness of the neutrino masses is due to a loop suppression and the additional particles carry a non-trivial charge under an unbroken symmetry which is responsible for DM stability. The simplest idea in this regard is the so-called Scotogenic model~\cite{Ma:2006km}, where the neutrino masses are generated at the 1-loop level. In this model, the DM candidate happens to be the lightest particle running inside the loop with an odd charge under a $\mathbb{Z}_2$ discrete symmetry. It could be either bosonic, a CP-even (odd) scalar, or fermionic, a heavy Majorana particle. The strong connection between DM and neutrino mass generation has driven novel studies within this context \cite{Hagedorn:2018spx,Blennow:2019fhy} as well as new variants \cite{FileviezPerez:2019cyn}.\\

Here we have considered the case where the neutrino mass is generated after the spontaneous breaking of lepton number in the Scotogenic model~\cite{Babu:2007sm} leading to the existence of the Majoron, $J$, a physical Nambu-Goldstone boson~\cite{Chikashige:1980ui, Schechter:1981cv}.
As a consequence, an invisible Higgs decay channel opens up contributing to its total decay width~\cite{Joshipura:1992ua, romao:1992zx,
joshipura:1992hp, Bonilla:2015uwa}. On top of that, in this model
there are two DM annihilation channels  when the DM is a Majorana fermion. One is mediated by $\mathbb{Z}_2$-odd
particles (t-channel)~\cite{Kubo:2006yx} and the other one (s-channel)~\cite{Babu:2007sm,Aranda:2018lif} coming from the mixing between the scalar singlet
and the SM model Higgs after  the spontaneous  breaking of lepton number and electroweak symmetries.
The latter helps to explain DM relic abundance in the Universe for DM masses below the TeV region. \\

We organized the paper as follows: we introduce the model in the next Section. All the constraints
used in our analysis are given in Section~\ref{sec:constraints}. We describe how
the analysis is made and we present our results in Section~\ref{sec:analysis}. Finally, we conclude in
Section~\ref{sec:conclusions}.

\section{The model}
\label{sec:Models}

We consider a model where a scalar singlet $\sigma$, a $SU(2)_L$ scalar doublet $\eta$
with hypercharge $1/2$, and three generations of Majorana fermions $N_i$ (with $i=1,2,3$)
are added to Standard Model. It is assumed that the scalar doublet $\eta=(\eta^+,\eta^0)^T$ and the Majorana
fermions have an odd charge under an unbroken discrete $\mathbb{Z}_2$ symmetry. This setup can be seen as an extension of the \emph{Scotogenic}
model~\cite{Ma:2006km}. Hence, the lightest $\mathbb{Z}_2$-odd particle turns out to be a stable DM candidate.
Furthermore, we consider the case where the masses of the heavy Majorana fermions are dynamically generated when the scalar singlet gets a vacuum expectation value $\langle \sigma \rangle $. This requires that the scalar singlet $\sigma$ has a non-trivial charge under lepton number and is responsible of the neutrino mass generation after spontaneous symmetry breaking. The particle content and charge assignments of the model are shown in Table~\ref{tab:Model}.

\begin{table}[t]
\centering
\begin{tabular}{|c|c|c|c|c|c|c|}
\hline \phantom{XXXXXXXX} & \phantom{X} $L_i$ \phantom{X} & \phantom{X}$\ell_{R_i}$\phantom{X} & \phantom{X} $\Phi$ \phantom{X} & \phantom{X} $\eta$ \phantom{X} & \phantom{X} $N_i$ \phantom{X} & \phantom{X} $\sigma$ \phantom{X} \\
\hline \hline
{\bf \color{blue} $SU(2)_L$} & {\bf \color{blue} $2$} & {\bf \color{blue} $1$} & {\bf \color{blue} $2$} & {\bf \color{blue} $2$} & {\bf \color{blue} $1$} & {\bf \color{blue} $1$} \\
\hline
{\bf \color{blue} $U(1)_Y$} & {\bf \color{blue} $-1/2$} & {\bf \color{blue} $-1$} & {\bf \color{blue} $1/2$} & {\bf \color{blue} $1/2$} & {\bf \color{blue} $0$} & {\bf \color{blue} $0$} \\
\hline
{\color{red} $U(1)_L$} & {\color{red}$-1$} & {\color{red}$-1$} & {\color{red}$0$} & {\color{red}$0$} & {\color{red}$-1$} & {\color{red}$2$}  \\
\hline
$\mathbb{Z}_2$ & $+$ & $+$ & $+$  & $-$ & $-$ & $+$ \\
\hline
\end{tabular}
\caption{Particle content and charge assignments of the model. \label{tab:Model}}
\end{table}

Considering the particle content and additional symmetries, the renormalizable $\mbox{SM}\,\otimes\, U(1)_L \otimes\mathbb{Z}_2$ invariant Lagrangian for leptons is given by:
\begin{equation}
-\mathcal{L}_{\text{Y}} \supset Y^{\ell}_{ij} \bar{L}_i \Phi \ell_{R_j} + Y^{\nu}_{ij} \, \bar{L}_i  \tilde{\eta} N_j + \cfrac{1}{2} Y^N_{ij} \, \sigma \bar{N}^c_i N_j + \textit{h.c.},
\label{eq:LagYL}
\end{equation}
where $\tilde{\eta} = i \tau_2 \eta^*$, $L_i=(\nu_{L_i}, \ell_{L_i})^T$ with $i,j= ~e,~\mu$ and $\tau$. The scalar
fields

\begin{eqnarray}
\Phi=
\begin{pmatrix}
\phi^+\\
 \phi^0
\end{pmatrix} & \ \ \mbox{and} &\ \ \eta=
\begin{pmatrix}
\eta^+\\
 \eta^0
\end{pmatrix},
\end{eqnarray}
denote the usual SM Higgs doublet and the inert doublet respectively. On the other hand, the scalar potential of the model reads
\begin{eqnarray}
\label{potential}
V &=  \mu_1^2 \Phi^\dagger \Phi + \mu_2^2 \eta^\dagger \eta + \mu_3^2 \sigma^* \sigma + \lambda_1 (\Phi^\dagger \Phi)^2 + \lambda_2 (\eta^\dagger \eta)^2 + \lambda_3 (\eta^\dagger \eta)(\Phi^\dagger \Phi) \nonumber\\
&+ \lambda_4 (\eta^\dagger \Phi) (\Phi^\dagger \eta ) + \displaystyle \frac{\lambda_5}{2} \left[ (\eta^\dagger \Phi)^2 + (\Phi^\dagger \eta)^2 \right] + \lambda_6 (\sigma^* \sigma)^2  \\
&+ \lambda_7 (\sigma^* \sigma) (\Phi^\dagger \Phi) + \lambda_8 (\sigma^* \sigma) (\eta^\dagger \eta).\nonumber
\end{eqnarray}
For simplicity, the dimensionless parameters $\lambda_i$ (with $i=1,...,8$) in the last equation are assumed
to be real. The scalar singlet $\sigma$ and the neutral component of the doublet $\Phi$
in eq.~(\ref{potential}) can be shifted as follows
\begin{equation}
\label{eq:shifts}
\sigma = \frac{v_\sigma}{\sqrt{2}} + \frac{R_1 + i\, I_1}{\sqrt{2}}
\ \ \text{and} \ \
\phi^0 = \frac{v_\Phi}{\sqrt{2}} + \frac{R_2 + i\, I_2}{\sqrt{2}},
\end{equation}
where $v_{a}$ (with $a=\sigma,\Phi$) are the vacuum expectation values and $v_\Phi=246$~GeV;
$R_j$ and $I_j$ (with $j=1,2$) represent the CP-even and CP-odd parts of the fields.

\subsection{Mass spectrum}
Computing the second derivatives of the scalar potential in eq.~(\ref{potential})
and evaluating them at the minimum of the potential, one gets the CP-even and CP-odd
mass matrices, $M^2_R$ and $M^2_I$ respectively.
There are two CP-odd massless fields,  one of them corresponds to the longitudinal component of the
$Z$ boson and the other one is a physical Nambu-Goldstone boson resulting from the spontaneous breaking
of the $U(1)_L$ symmetry, the Majoron $J$~\cite{Chikashige:1980ui,Schechter:1981cv}. Hence,
\begin{equation}
  \label{eq:goldstons}
  J\equiv I_1, \ \ G^0 \equiv I_2\ .
\end{equation}

For the CP-even part, one can define the two mass eigenstates $h_i$ through the rotation matrix $O_R$ as follows,
\begin{equation}
  \label{eq:rotation}
  \begin{pmatrix}
    h_1\\[+2mm]
    h_2
  \end{pmatrix}
=
O_R
\begin{pmatrix}
  R_1\\[+2mm]
R_2
\end{pmatrix}
\equiv
\begin{pmatrix}
  \cos\alpha & \sin\alpha\\[+2mm]
-\sin\alpha &\cos\alpha
\end{pmatrix}
\
\begin{pmatrix}
  R_1\\[+2mm]
R_2
\end{pmatrix}\, .
\end{equation}
The angle $\alpha$ is interpreted as the doublet-singlet mixing angle. Then, we have that
\begin{equation}
  \label{eq:Hmasses}
  O_R\, M^2_R\, O^T_R = \text{diag}(m^2_{h_1},m^2_{h_2})\, .
\end{equation}
where $M_R^2$ is the squared CP-even mass matrix whose eigenvalues are given by,
\begin{equation}
\label{eq:mh12}
m^2_{(h_1,h_2)} = \left(\lambda_1 v_\Phi^2  + \lambda_6 v_\sigma^2 \right) \mp \sqrt{\lambda^2_7 v_\Phi^2 v_\sigma^2 +(\lambda_1 v_\Phi^2 - \lambda_6 v_\sigma^2)^2},
\end{equation}
where the ``$-$'' (``$+$'') sign corresponds to $h_1$ ($h_2$). Notice that one of these scalar has
to be associated to the SM Higgs boson with a $125.09$~GeV mass~\cite{Aad:2015zhl}.
Furthermore, the masses of the CP-even and CP-odd components of the inert doublet, $\eta$,
turn out to be
\begin{equation}
m^2_{(\eta_R,\eta_I)} = \mu^2_2+\frac{\lambda_8}{2}v_\sigma^2 + \frac{\lambda_3+\lambda_4 \pm \lambda_5}{2}v_\Phi^2.
\end{equation}
The mass of the charged scalar field is given by,
\begin{equation}
m^2_{\eta^\pm} = \mu_2^2 + \frac{\lambda_3}{2}v_\Phi^2 + \frac{\lambda_8}{2}v_\sigma^2.
\end{equation}
Notice that the masses of the CP-even and CP-odd fields satisfy the relation $ \lambda_5 v_\Phi^2 = (m^2_{\eta_R} - m^2_{\eta_I})$.\\

\begin{figure}[t]
  \centering
  \includegraphics[width=0.4\textwidth]{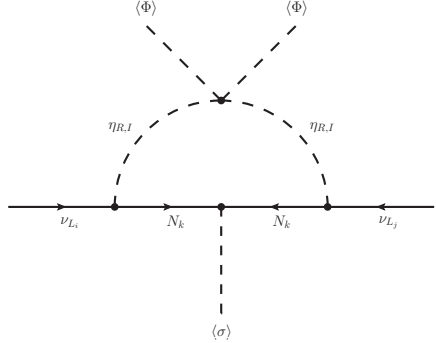}
  \caption{One-loop Feynman diagram for neutrino mass generation.}
  \label{fig:loop}
\end{figure}

As it was mentioned before, neutrino masses are generated dynamically like the rest of the SM fermions.
That is, the Majorana masses of $N_i$ as well as the light neutrinos arise after the spontaneous breaking
of the global $U(1)_L$ symmetry. From eq.~(\ref{eq:LagYL}) follows that the mass matrix for the $N_i$ fields is given by
\begin{equation}
(m_{N})_{ij} = \sqrt{2} Y^N_{ij} v_\sigma .
\label{eq.RHNmass}
\end{equation}
The one-loop neutrino mass generation is depicted in Fig.~\ref{fig:loop}.
After the electroweak symmetry breaking one gets that the light neutrino mass matrix is
given by the following expression ~\cite{Ma:2006km,Babu:2007sm}
\begin{equation}
\label{eq:numass}
(M_\nu)_{ij}
= \sum_{k=1}^3 \frac{Y^\nu_{ik} \, Y^\nu_{kj} m_{N_k}}{32 \pi^2}\left[ \frac{m_{\eta_R}^2}{m_{\eta_R}^2-m_{N_k}^2} \log \frac{m_{\eta_R}^2}{m_{N_k}^2} -  \frac{m_{\eta_I}^2}{m_{\eta_I}^2 - m_{N_k}^2} \log \frac{m_{\eta_I}^2}{m_{N_k}^2} \right].
\end{equation}
%
\section{Summary of constraints}
\label{sec:constraints}

Before analysing the sensitivities of the experimental searches for WIMPs, we first discuss the theoretical and experimental restrictions that are implemented in our analysis.

\subsection{Boundedness conditions}
In order to ensure that the theory is perturbative the quartic couplings in the scalar potential, eq.~(\ref{potential}), as well as the Yukawa couplings in eq.~(\ref{eq:LagYL}) are limited to be~\cite{Lindner:2016kqk},
\begin{equation}
|\lambda_i|, |Y^{a}_{jk}|^2 \leq 4 \pi\ \ \text{with} \ \ i=1,...,8;\, j,k=1,2,3\ \ \text{and} \ \ a=\nu,\ell, N.
\end{equation}
Furthermore, the consistency requirements of the scalar potential demand that the dimensionless parameters in eq.~(\ref{potential}) have to fulfill the following conditions~\cite{Kadastik:2009cu},
\begin{eqnarray}
\centering
& & \lambda_1,\lambda_2,\lambda_6 \geq 0, \ \
\lambda_3 \geq -2\sqrt{\lambda_1\lambda_2}, \nonumber \\
& & 4 \lambda_1\, \lambda_6 \geq \lambda_{7}^2,\ \
4 \lambda_2\, \lambda_6 \geq \lambda_{8}^2 \ \ \text{and}\ \
\lambda_3+\lambda_4-|\lambda_5|\geq -2\sqrt{\lambda_1\lambda_2}.
\end{eqnarray}
From the last relations it is guaranteed that the scalar potential is bounded from below.

\subsection{Searches of new physics}

As we described in the previous section, there are 6 physical scalars in the model: three CP-even $h_i$ ($i=1,2$) and $\eta_{R}$; two CP-odd $\eta_I$ and the Majoron $J$; and a charged scalar $\eta^\pm$. Therefore, one has to impose the constraints on the scalar masses coming from the LEP results~\cite{Heister:2002ev} and the latest reports from the
LHC on the Higgs properties~\cite{Tanabashi:2018oca}.
Notice that the invisible Higgs decay channel is always present, namely the Higgs decay
into Majorons $h\to JJ$, where in our case the SM Higgs $h$ will be identified with either $h_1$ or $h_2$. Then, this decay
mode coexists with the Higgs decay into the fermion dark matter, $N_1$, when it is kinematically allowed, i.e. $h\to N_1 N_1$ when $m_{N_1} < m_{h}/2$.
Therefore, we consider~\cite{Tanabashi:2018oca}
\begin{equation}
 \mathcal{B}_{inv}\equiv \text{BR}(h\to \text{invisible}) < 0.28\ \ \text{at} \ \ 95\% \ \ \text{C.L}.
\end{equation}
On the other hand, the LEP collaboration studies on the invisible decays of $W^\pm$ and $Z^0$ gauge bosons~\cite{Heister:2002ev} provide bounds on the masses of the inert scalars $\eta_{R}(\eta_I)$ and $\eta^\pm$. From these searches, the following conditions must be fulfilled~\cite{Lundstrom:2008ai}
\begin{equation}
m_{\eta_R} + m_{\eta_I}>m_Z, \ \
m_{\eta^\pm }> m_{Z}/2\ \ \text{and} \ \  m_{\eta^\pm} + m_{\eta_{(R,I)}}> m_W.
\end{equation}
The LEP reports also established disallowed mass regions for the mass splitting given by,
\begin{equation}
 m_{\eta_I} - m_{\eta_R} > 8~\text{GeV}\ \ \text{if} \ \ m_{\eta_R}< 80~\text{GeV} \ \ \text{and} \ \ m_{\eta_I}< 100~\text{GeV},
\end{equation}
and  $m_{\eta^\pm}>80$~GeV.\\

Finally, it is important to mention that the oblique parameters $S$, $T$ and $U$ are also sensitive to new physics~\cite{Peskin:1990zt,Peskin:1991sw}.
Then, it has to be considered that values of these parameters in the model lie within the following regions~\cite{Tanabashi:2018oca}.

\begin{equation}
\label{eq:stu}
S=0.02\pm0.10,  \quad T=0.07\pm 0.12 \quad \text{and} \quad U=0.0\pm0.09.
\end{equation}

\subsection{Dark matter searches}

The abundance of DM in the Universe, given in terms of the cosmological abundance parameter, eq.~(\ref{eq:omega}), provides restrictions on the parameter space of DM models.
Furthermore, there exist constraints coming from searches of DM by experiments using (in)direct detection
techniques. The direct dark matter detection
experiments have set bounds, for DM masses above 6 GeV, on the dark matter-nucleon
spin-independent scattering cross section. The most stringent bounds are set by the XENON1T experiment,
that is $ \sigma_{\text{SI}} \lesssim 4.1\times10^{-47} \text{cm}^2$ for a DM mass of 30 GeV at 90\% C.L ~\cite{Aprile:2018dbl}.
On the other hand, the astronomical gamma ray observations constrain the velocity averaged cross section of dark matter
annihilation into gamma rays $\langle \sigma v \rangle_\gamma$. The Fermi-LAT satellite has performed this indirect DM search and constraint the
cross section to be $\langle \sigma v \rangle_\gamma \lesssim 10^{-29} \text{cm}^{3}\text{s}^{-1}$ ~\cite{Ackermann:2013uma}.
Notice that there are promising searches using neutrino telescopes like IceCube~\cite{Halzen:2010yj}, Antares~\cite{Collaboration:2011nsa},
and KM3Net~\cite{Adrian-Martinez:2016fdl}. Limits on the annihilation cross section for the typical WIMP mass range are not as competitive as other limits
obtained with other astroparticle messengers. In additon, neutrinos are used to set bounds on spin-dependent direct dectection cross section infered from the capture and annihilation of DM in the Sun~\cite{Aartsen:2016exj, Bhattacharya:2019ucd}. The tightest limits are at $m_{DM} \simeq 500$~GeV with $\sigma_{SD} \lesssim 10^{-40}\, {\rm cm}^2$..

\subsection{Neutrino oscillation parameters}
The neutrino masses are obtained after diagonalisation of the mass matrix given in eq.~(\ref{eq:numass}).
The relation between $M_\nu$ and the diagonal mass matrix is given by,
\begin{equation}
\label{eq:leptonmixing}
 M_\nu= U_\nu^* \, \text{diag}(m_{\nu_1}, m_{\nu_2}, m_{\nu_3}) \,U_\nu^{\dagger}
\end{equation}
where $m_{\nu_i}$ are the neutrino masses. Here we are not assuming a flavor-diagonal mass matrix for chaged leptons. Therefore, the lepton mixing matrix is difined as $U_{L}\equiv U_\ell^\dagger U_\nu= U_L (\theta_{12}, \theta_{13}, \theta_{23},\delta_{\text{CP}})$, where $\theta_{ij}$ are the mixing angles and $\delta_{\text{CP}}$ corresponds to the Dirac CP-violating phase.
$U_\nu$ and $U_\ell$ are the matrices that diagonalise the neutrino and charged lepton square mass matrices $M_\nu^\dagger M_\nu$ and $M_\ell M_\ell^\dagger$ respectively. The lepton mixing angles
$\theta_{ij}$ are determined by neutrino oscillation experiments. From global fits of neutrino
oscillation  parameters~\cite{deSalas:2017kay} (for other fits of neutrino
oscillation  parameters we refer the reader to \cite{Capozzi:2016rtj,Esteban:2018azc}) the
best fit values  and the $1\sigma$ intervals for a normal neutrino mass ordering (NO) are

\begin{eqnarray}
& &|\Delta m_\text{sol}^2|= 7.55^{+0.20}_{-0.16}\, \times 10^{-5}\,\text{eV}^2,\ \ |\Delta m_\text{atm}^2|= 2.50\pm 0.03\, \times 10^{-3}\,\text{eV}^2, \notag \\
& & \theta_{12} / ^{\circ} =34.5^{+1.2}_{-1.0}, \ \
\theta_{13} / ^{\circ} =8.45^{+0.16}_{-0.14}, \ \ \theta_{23} / ^{\circ} =47.7^{+1.2}_{-1.7},
\ \ \text{and}\ \ \delta_{\text{CP}} / ^{\circ} =218^{+38}_{-27}. \ \
\end{eqnarray}


\section{Numerical analysis}
\label{sec:analysis}
We have mentioned that the nature of the DM candidate in this model could be either
fermionic or scalar. This is the lightest particle with odd charge under the $\mathbb{Z}_2$ symmetry
and running in the neutrino mass generation loop as shown in Fig.~\ref{fig:loop}.\\

\begin{figure}[t]
\centering
\includegraphics[width=0.25\textwidth]{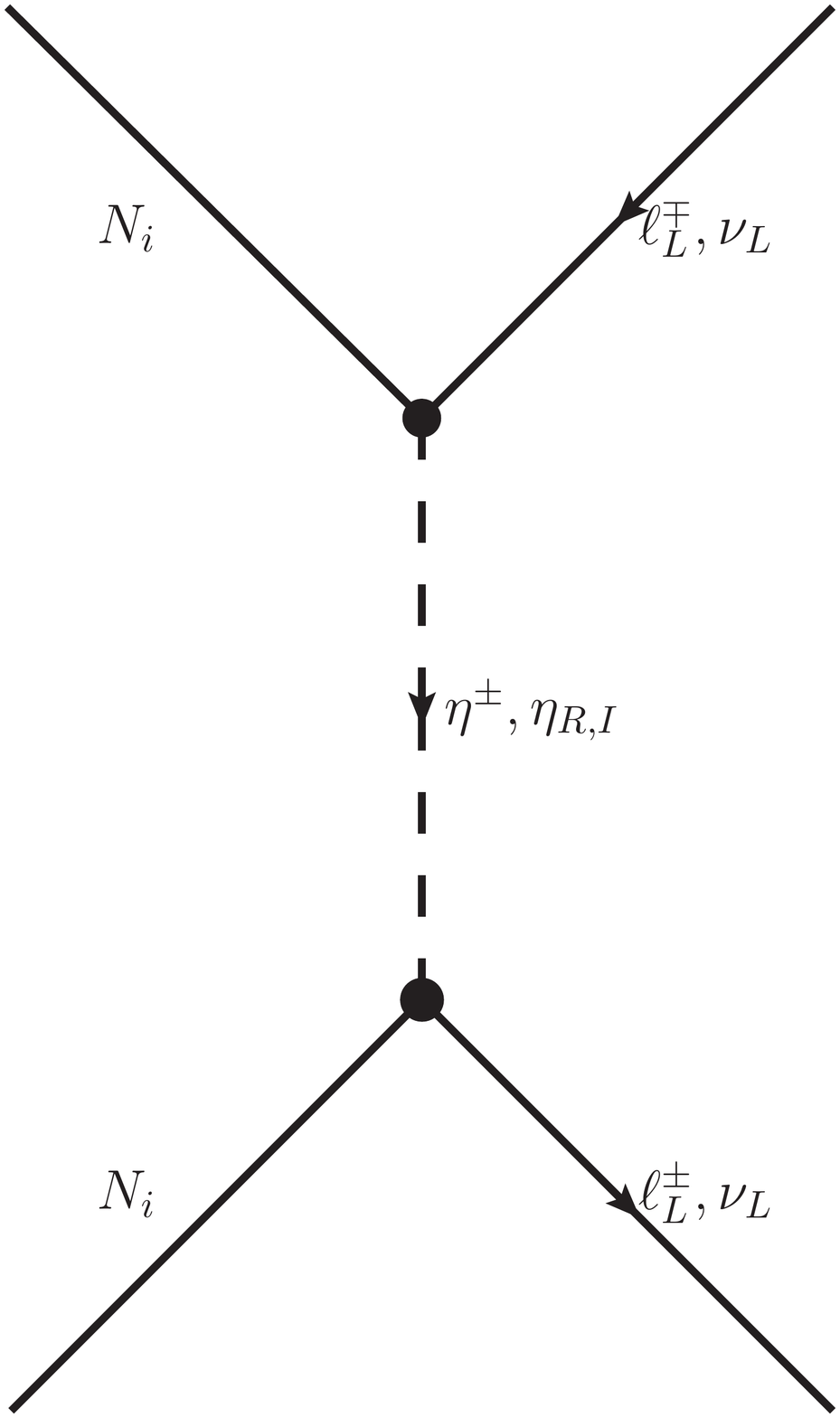}\ \ \ \ \ \ \ \
\includegraphics[width=0.25\textwidth]{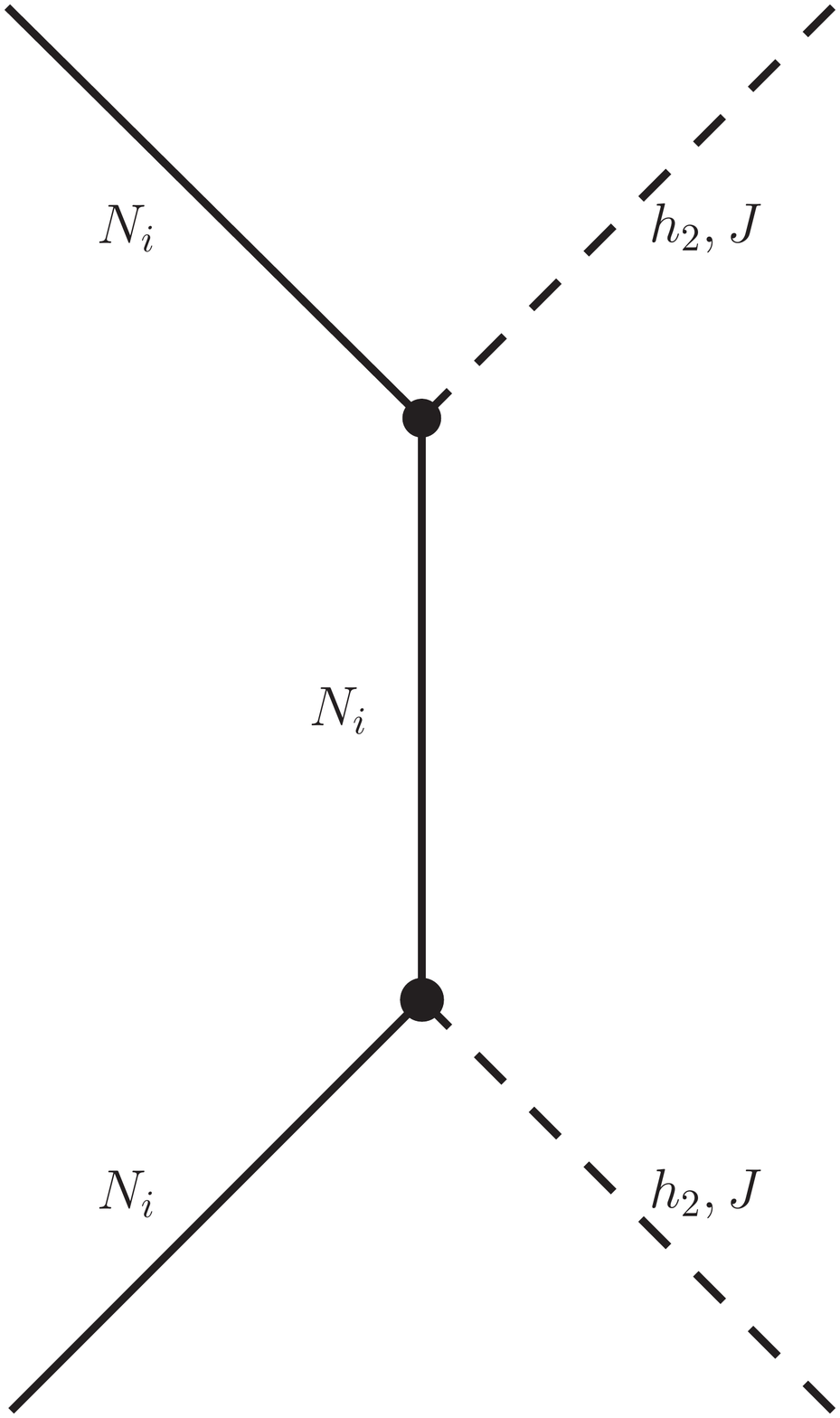}\\
\includegraphics[width=0.33\textwidth]{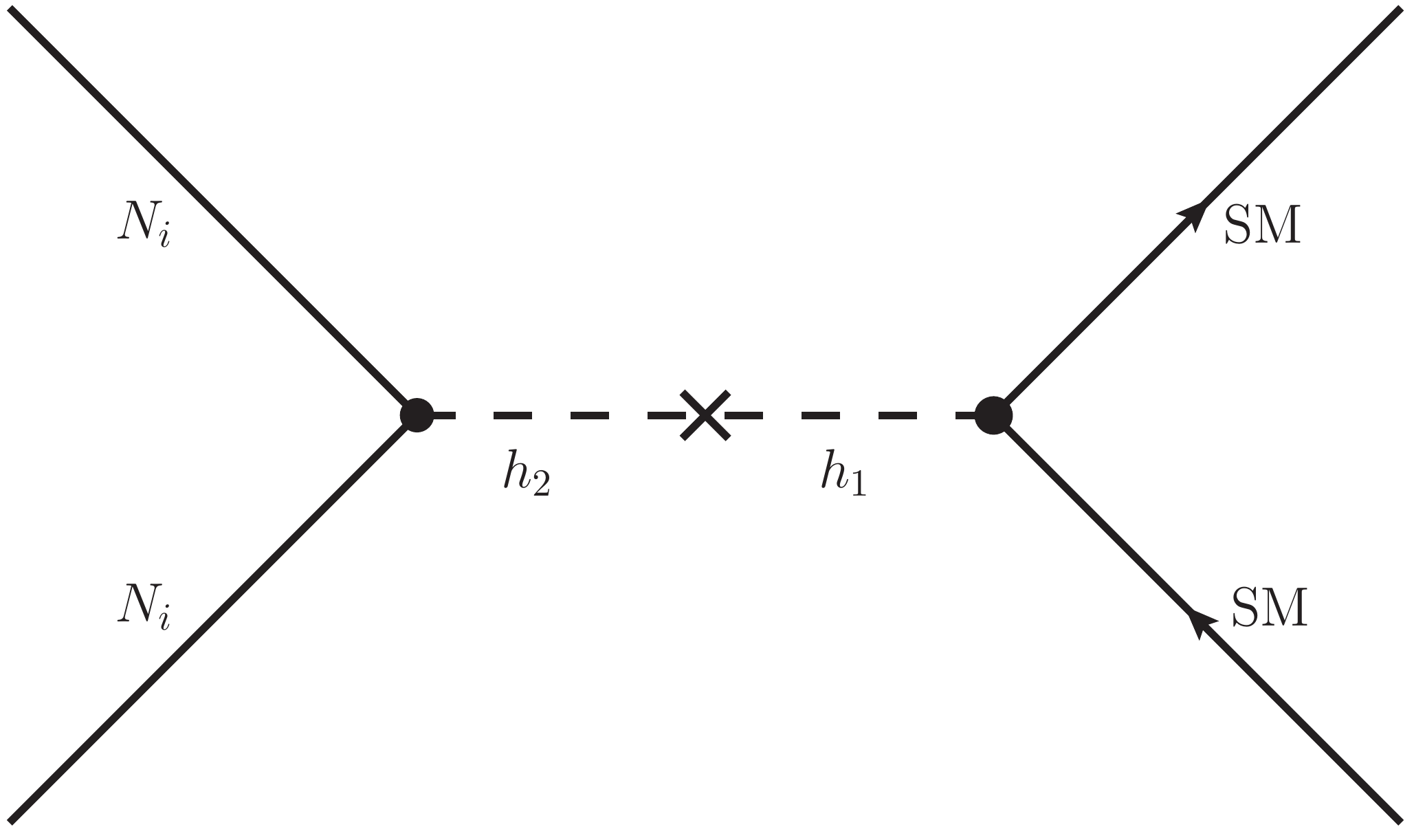}
\includegraphics[width=0.33\textwidth]{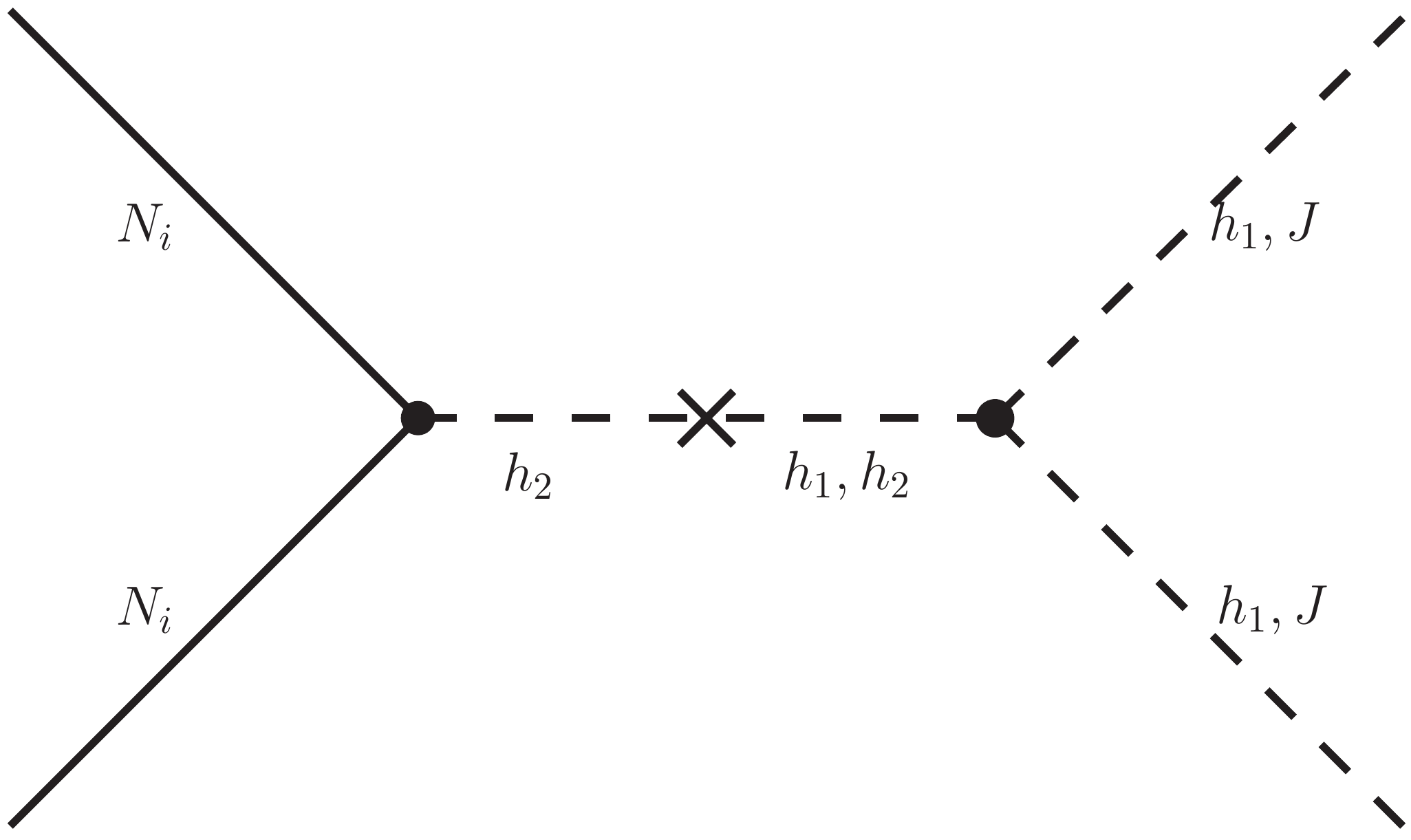}
\caption{Feynman diagrams for the annihilation channels of the fermion dark matter in the model. On the left annihilation into SM particles. On the right annihilation into Majorons and higgses.}
\label{fig:tschannels}
\end{figure}

In our study we will focus in the case in which the DM is the lightest Majorana particle \footnote{
In Ref.~\cite{Toma:2013zsa,Vicente:2014wga,Ibarra:2016dlb} has bee analyzed the situation where the DM
is the Majorana fermion within the simplest scotogenic model. Note that the case in which
the DM is the neutral component of the inert doublet $\eta$ is similar to the studies for inert doublet model~\cite{Belyaev:2016lok}.}, i.e. $N_1$.
Therefore, in this case the DM annihilates via the t- and s-channel In Fig.~\ref{fig:tschannels} . The former is mediated
by a Majorana fermion $N_i$ and by the inert scalars. It has been shown that the bounds on lepton flavor
violation (LFV) processes, e.g. $\mu\to e\gamma$,
demand small neutrino Yukawas (namely, $Y^\nu\ll1$) and hence the t-channel mediated by the inert scalars
becomes suppressed inducing DM overabundance~\cite{Toma:2013zsa,Vicente:2014wga}. However,
we show that we can keep suppressed the inert scalar mediated t-channel and thanks to
the s-channel mediated by the singlet $\sigma$ it is possible to account for the right amount
of DM relic abundance. As a result, it is crucial to have a non-vanishing mixing angle between
CP-even parts of the Higgs doublet $\Phi$ and the iso-singlet $\sigma$, eq.~(\ref{eq:rotation}),
and in agreement with current experimental data.

For the numerical analysis we have used the MicrOMEGAS~\cite{Belanger:2001fz} and performed a scan
over all free parameters of the model. For the dimensionless parameters in the scalar sector
we took the following intervals:
\begin{equation}
10^{-6}\leq |\lambda_{2,...,8}| \leq 1,
\end{equation}
and $\lambda_1$ being determined by the SM Higgs mass. We are taking $h_1$ as the SM Higgs
then  $m_{h_1}=125$~GeV. On the other hand, we are varying the mass of $h_2$ within the range $m_{h_2}\in[20,2000]$~GeV.
Notice that we are taking masses below 125~GeV which is in perfect agreement with both
the LHC and LEP constraints as long as the doublet-singlet mixing given by $\sin\alpha$ in
eq.~(\ref{eq:rotation}) is less than 20\% \cite{Bonilla:2015uwa}.

For the masses of the inert scalars we considered the following ranges,
\begin{equation}
m_{\eta_R}\in[110,5000]~\text{GeV}, \ \ m_{\eta^\pm}\in[135,5000]~\text{GeV},
\end{equation}
and mass of the CP-odd part $\eta_I$ is determined by using the
relation $ \lambda_5 v_\Phi^2 = (m^2_{\eta_R} - m^2_{\eta_I})$.
For the lepton number breaking scale, namely the singlet's vev $v_\sigma$, we have
used $v_{\sigma}\in[500,10000]~\text{GeV}$. Bear in mind that this vev
provides the mass of the heavy Majorana fermions, $N_i$, whose masses (taken to be diagonal)
are varied in the following ranges,
\begin{equation}
\label{eq:Nmasses}
m_{N_1}\in[8,1000]~\text{GeV} \ \ \text{and} \ \ m_{N_{2,3}}\in[100,5000]~\text{GeV}.
\end{equation}
Since $N_1$\footnote{There are regions of the parameter space for $M_{N_1}<8$ GeV but those are below the neutrino floor.} is the DM candidate of the theory we have to impose
$m_{N_1}<m_{N_{(2,3)}}<m_{\eta_{(R,I,\pm)}}$~\footnote{We discard contributions to the DM abundance
produced by co-annihilation processes, i.e. $m_{N_{(2,3)}}\lesssim 1.1\,m_{N_1}$.
}.
The above considerations are made in such a way that they all satisfy the theoretical and experimental
constraints described in Sec.~\ref{sec:constraints}. We computed the value of the $S$ and $T$ parameters
using the expressions given in~\cite{Grimus:2007if,Grimus:2008nb}, taking $U=0$~\cite{Tanabashi:2018oca} and keeping
those solutions that are in agreement with the bounds given in eq.~(\ref{eq:stu}). It is worth to mention that
we considered only S and T within  the $90\%$ level shown in Fig. (10.6) from reference~\cite{Tanabashi:2018oca}.
In addition, we calculated the light neutrino masses feeding the neutrino mass expression given in
eq.~(\ref{eq:numass}) and assumed normal ordering for neutrino masses~\footnote{For simplicity, we have assumed
that the Yukawa matrices $Y^{\nu}$ and $Y^N$ are real and diagonal. Following this assumption, the Yukawa matrix for charge
lepton is non-diagonal (such that, $U_{L}= U_\ell^\dagger U_\nu=U_\ell^\dagger$) in order to fit neutrino
oscillation experimental data in eq.~(\ref{eq:leptonmixing}).}. Then,
we took as valid only the points that satisfy the best fit values from the
global fit of neutrino oscillation parameters\footnote{The sum of neutrino masses was restricted
using the cosmological limit provided by Planck, namely $\sum m_\nu < 0.12$~eV \cite{Vagnozzi:2017ovm,Aghanim:2018eyx}.}~\cite{deSalas:2017kay}.\\

Our last requirement is that the annihilation cross section of the fermion DM candidate
into Majorons (see Fig.~\ref{fig:tschannels}) is subdominant at the moment of the freeze-out in order to guarantee detectability in DM direct detection experiments and to avoid direct detection cross sections in regions far below the neutrino floor.\\


\subsection{Viable dark matter mass regions}
\label{subs:results}

\begin{figure}[t]
\centering
\includegraphics[width=0.7\textwidth]{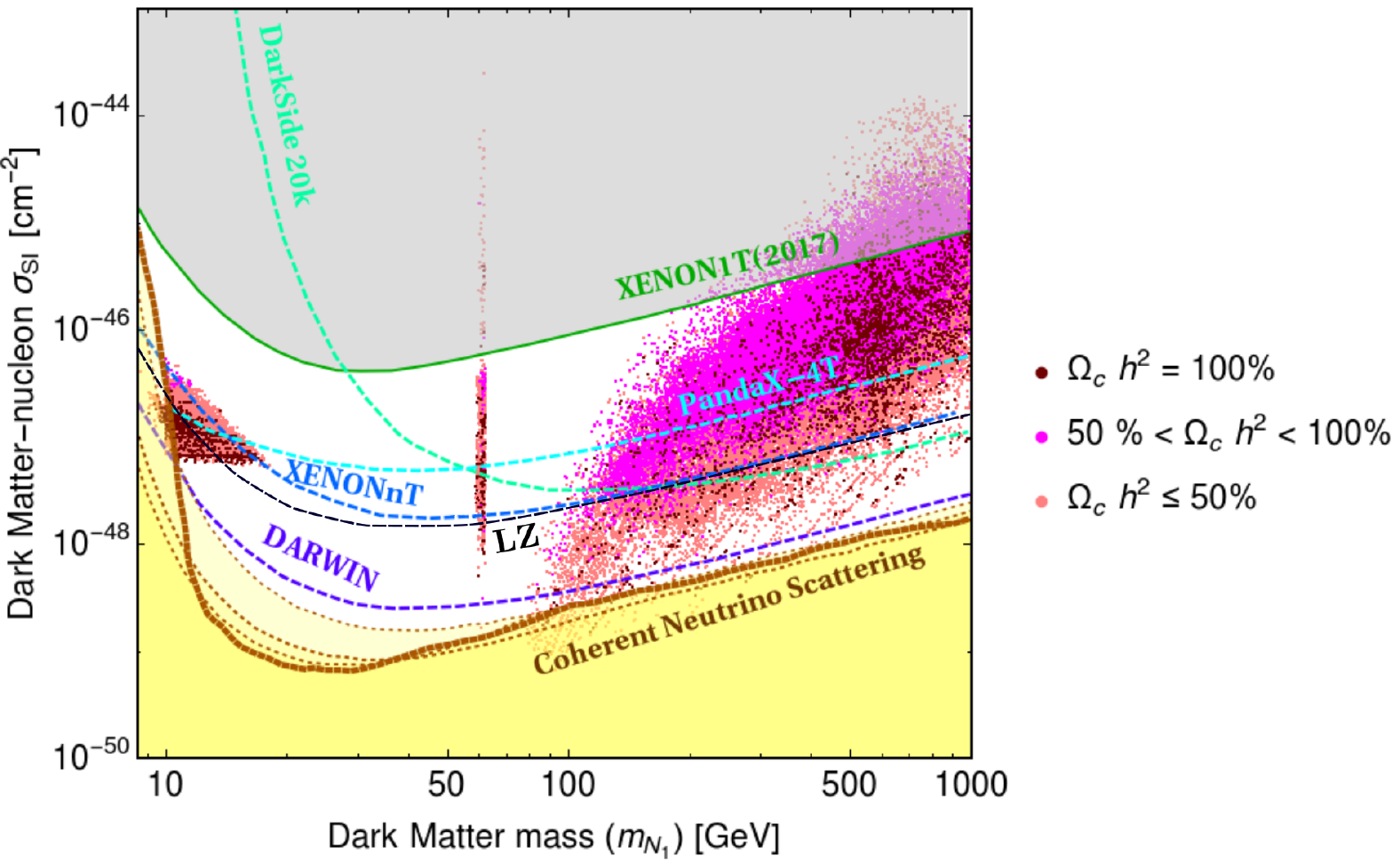}
\caption{$\text{Dark matter mass}-\sigma_{\text{SI}}$ plane showing the solutions in the model that satisfy all theoretical and experimental constraints given in Section~\ref{sec:constraints}. The latest bound on direct dark matter detection is set by the XENON1T experiment~\cite{Aprile:2018dbl} (top shaded area). The dashed lines represent the expected sensitivities in forthcoming experimental searches such as XENONnT~\cite{Aprile:2015uzo}, LUX-ZEPLIN (LZ)~\cite{Akerib:2018dfk}, DarkSide 20k~\cite{Aalseth:2017fik}, DARWIN~\cite{Aalbers:2016jon} and PandaX-4T~\cite{Zhang:2018xdp}.}
\label{Fig1}
\end{figure}

Following the considerations that we stated previously, we show in Fig.~\ref{Fig1}
the nucleon-dark matter spin-independent cross section $\sigma_{SI}$ as a function of the
fermion DM mass, $m_{N_1}$. From the numerical analysis we have found three different
viable mass regions for a fermion DM candidate within the model. We refer as viable to those solutions that fulfil the theoretical and experimental bounds given in Section~\ref{sec:constraints}.
These are:
\begin{itemize}
 \item {\bf \it the low mass region}, with an approximate DM mass range $8~\text{GeV}\lesssim m_{N_1}\lesssim 20~\text{GeV}$ and $b\bar{b}$
 as dominant annihilation channel;
 \item {\bf \it the resonant region}, where $m_{N_1}\lesssim m_h/2$ (with $m_h=125.09~\text{GeV}$); and
 \item {\bf \it the high mass region}, for DM masses above 80 GeV where the fermion DM
 annihilates efficiently into the gauge bosons, i.e. $N_1 N_1\to VV$ with $V=(Z,W)$.
\end{itemize}
In all these domains, the DM annihilation into Majorons $(N_1 N_1\to JJ)$ at the moment of the freeze-out is always below 10\%.
The latest bound coming from direct detection searches of dark matter particles is set the XENON1T experiment~\cite{Aprile:2018dbl} and is defined by the top shaded area in Fig.~\ref{Fig1}.
The dark red points showed in the $m_{N_1}-\sigma_{\text{SI}}$ plane account for 100\% of the DM relic abundance while the solutions in purple and pink correspond only to a fraction of the DM abundance.
Notice that in {\it the high mass region} it is most likely a fermion DM with a mass around 500~GeV accounting for the whole amount of DM in the Universe. There are few points around $m_{N_1}\sim 10$~GeV and $m_{N_1}\sim 100$~GeV that could not be distinguished from the neutrino floor background (bottom shaded area).
Fig.~\ref{Fig1} also displays the future sensitivities for dark matter searches in direct detection experiments such as XENONnT \cite{Aprile:2015uzo} and LUX-ZEPLIN~\cite{Akerib:2018dfk}, DarkSide-20k \cite{Aalseth:2017fik}, DARWIN \cite{Aalbers:2016jon}, and PandaX-4T \cite{Zhang:2018xdp}.
For completeness we provide three benchmarks in Appendix~\ref{app:benchmarks} within each mass neighbourhood and their corresponding outputs.

\begin{figure}[t]
\centering
\includegraphics[width=0.7\textwidth]{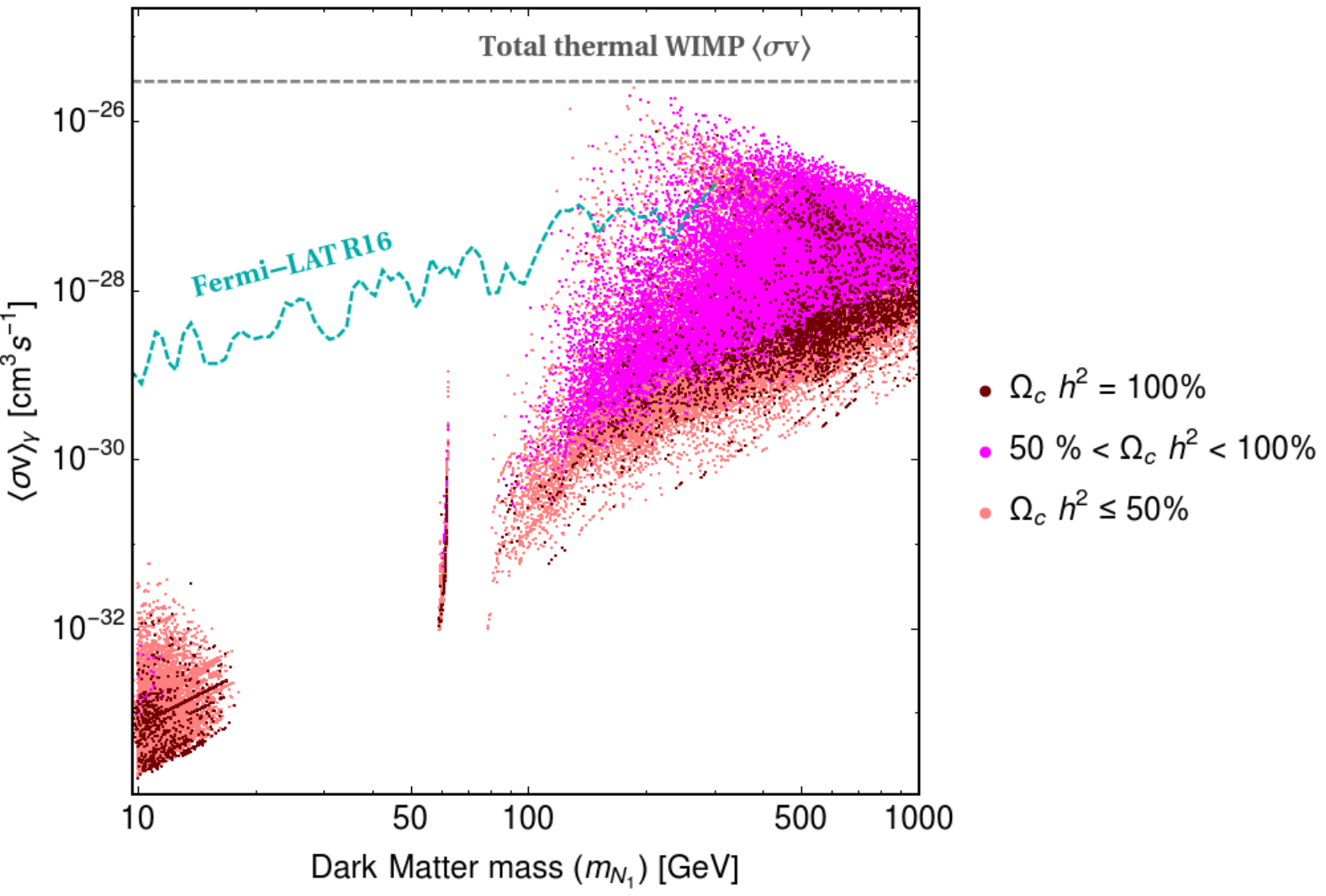}
\caption{Predictions for the velocity averaged cross section of dark matter
annihilation into gamma rays $\langle \sigma v \rangle_\gamma$ as function of the dark matter mass $m_{N_1}$. The dashed line represent the limit set by Fermi-LAT satellite results ~\cite{Ackermann:2013uma}. }
\label{Fig2}
\end{figure}

Fig.~\ref{Fig2} shows the predictions for the velocity averaged cross section of dark matter
annihilation into gamma rays $\langle \sigma v \rangle_\gamma$ as function of dark matter mass $m_{N_1}$.
We have found that the annihilation cross section of dark matter into gamma rays is up to two orders magnitude below the limit set by Fermi-LAT satellite results ~\cite{Ackermann:2013uma} on the indirect DM search (cyan dashed line in Fig.~\ref{Fig2}). This is the case for fermion DM with a mass inside the {\it low mass region}.
One can see that there are solutions in the {\it high mass region} that are ruled out by observations. In particular, the indirect search of DM excludes some points where the fermion DM represent only a fraction of the DM relic abundance. As before, all points satisfy the theoretical and experimental constraints listed in Section~\ref{sec:constraints} and the dark red points correspond to the solutions that account for the whole amount of DM in the Universe. The lighter (purple and pink) colors would require the existence of other DM candidates to explain observations, eq.~(\ref{eq:omega}).

\section{Conclusions}
\label{sec:conclusions}
In this work we have studied the scotogenic model with spontaneous breaking of lepton number. We have shown that it is possible to account for the whole amount of DM relic density thanks to the scalar singlet used to break lepton number which mixes with the CP-even part of the SM Higgs doublet. Notice that this DM annihilation portal  is absent in the simplest version of the scotogenic model, where lepton number is explicitly broken by the Majorana mass term, $N_i N_i$. In our analysis the LFV processes are suppressed because the neutrino Yukawas are kept small, then experimental constraints are naturally respected. We present a numerical analysis of the parameter space of the model and the predictions for the nucleon-dark matter spin-independent cross section $\sigma_{SI}$. We show that there are three different DM mass regions that can explain the DM relic abundance, satisfy current experimental constraints as well as the limits on $\sigma_{SI}$ reported XENON1T. We also included the future sensitivities of experiments that are devoted to search for the direct dark matter detection.


\begin{acknowledgments}
We would like to thank A. Vicente and one of the referees for pointing out a missing factor $1/2$ in eq.~\ref{eq:numass}.
The work of C.B. was supported by the Collaborative Research Center SFB1258. R.L. was supported by
Universidad Católica del Norte through the Publication Incentive program No. CPIP20180343.
LMGDLV, J.M.L. and E.P.  are supported  by DGAPA-PAPIIT IN107118 and CONACyT CB-2017-2018/A1-S-13051 (M\'exico). C.B. would like to thank IFUNAM for the hospitality while part of this work was carried out.
\end{acknowledgments}

\appendix

\section{Benchmarks}
\label{app:benchmarks}

Here we present three benchmarks (BM1, BM2, BM3) corresponding to the different mass regions described in Section~\ref{subs:results}
where the fermion DM satisfy all experimental and theoretical constraints summarized in Section~\ref{sec:constraints},
see Tables~\ref{tab:BM-I} and~\ref{tab:BM-II}. Additionally, these representative points are such that the DM particle $N_1$
constitute 100\% of the relic abundance in the Universe, see Table~\ref{tab:BM-III}.

The values for the dimensionless parameters in the Lagrangian as well as the dimensionful parameters in the scalar
potential are shown in Table~\ref{tab:BM-I}. Notice that the neutrino Yukawas $Y^\nu_i$ (with $i=1,2,3$) are
small and as a result LFV processes are suppressed. We include as example in Table~\ref{tab:BM-II} the
branching fractions of $\mu\to e\gamma$ for each benchmark.

\begin{table}[t]
	\centering
	\begin{tabular}{|c|c|c|c|c|c|c|c|c|c| }
	\hline
		& $m_{N_1}$[GeV] & $m_{N_2}$[GeV] & $m_{N_3}$[GeV] & $m_{\eta_R}$[GeV] & 	$m_{\eta_I}$[GeV] & $m_{\eta^\pm}$  [GeV] & $m_{h_2}$[GeV] & $v_\sigma$[TeV] & $\mu_2^2$  [GeV${}^2$] 
		\\
		\hline
		\hline
		BM1 & 10 & 119 & 316 & 520 & 536 & 486 & 20.7 & 10.4&                2.1$\times10^{5}$           \\
		\hline
		BM2 & 59.1 & 184 & 410 & 666 & 675 & 645 & 149 & 1.08 &          4.60$\times10^{5}$           \\
		\hline
		BM3 & 707 & 924 & 940  & 1132 & 1119 & 1119 & 1498 & 5.80 & 			 6.98$\times10^{6}$\\
		\hline
	\end{tabular}

	\begin{tabular}{|c|c|c|c|c|c|c|c|c|}
	\hline
		&  $\sin 		\alpha$    & $\lambda_1$ & $\lambda_2$& $\lambda_3$           & $\lambda_4$ & $\lambda_5$ &$\lambda_6$& $\lambda_8$\\
		\hline
		\hline
		BM1 & 1.31$		\times10^{-1}$ & 1.27$\times10^{-1}$ & 8.93$\times10^{-1} $ & -6.7$	\times10^{-1}$ & 	1.4              & -2.82$\times10^{-1}$&3.17$\times10^{-6}$ & 8.19$\times10^{-4}$      \\
		\hline
		BM2 & 1.57$	  	\times10^{-1}$ & 1.30$\times10^{-1}$ & 1                     & -1.4$	\times10^{-1}$ & 	1.1              & -1.99$\times10^{-1}$&9.35$\times10^{-3}$ &-6.76$\times10^{-2}$ \\
		\hline
		BM3 &1.63$		\times10^{-1}$ & 9.23 $\times10^{-1}$ & 1.63    & 	1.63              &6.47 $\times10^{-1}$&-3.40$\times10^{-1}$ &3.24$\times10^{-2}$& -1.48$\times10^{-1}$\\
		\hline
	\end{tabular}

	\begin{tabular}{|c|c|c|c|c|c|c|}
	\hline
		&  $Y^N_1$ &$Y^N_2$ &$Y^N_3$ & $Y^\nu_1$ & $Y^\nu_2$  & $Y^\nu_3$\\
		\hline
		\hline
		BM1 &6.78$	\times10^{-4}$ &8.04$	\times10^{-3}$ &2.14$	\times10^{-2}$ & -1.61$		\times10^{-4}$ &-4.99$\times10^{-5}$ &-4.18$\times10^{-5} $\\
		\hline
		BM2 &3.86$	\times10^{-2}$ &1.20$	\times10^{-1}$ &2.68$	\times10^{-1}$ &-3.22$	  	\times10^{-5}$ &-2.75$\times10^{-5}$ & -4.70         $	\times10^{-5}$          \\
		\hline
		BM3 &8.61$	\times10^{-2}$ &1.12$	\times10^{-1}$ &1.14$	\times10^{-1}$ & -6.33$		\times10^{-6}$ & -8.88$\times10^{-5}$ & -3.17$\times10^{-5}$\\
		\hline
	\end{tabular}
	\caption{The values for the dimensionful parameters are shown on the top table. The values for the dimensionless parameters in the scalar sector and the neutrino sector are also given.
	}
	\label{tab:BM-I}
\end{table}

\begin{table}[t]
	\centering
	\begin{tabular}{|c|c|c|c|c|}
	\hline
		& $\mathcal{B}_{\text{inv}}$ &BR($ h_2 \to  J J  $) & BR($ h_2 \to  h_1 h_1  $) &   $\Gamma(h_1)$[MeV] \\
		\hline
		\hline
		BM1 &3.5$	\times10^{-2}$ &9.8$	\times10^{-2}$ & -                     & 3.9 \\
		\hline
		BM2 & 9.7$	\times10^{-2}$ &9.2$	\times10^{-1}$ & -                     & 4.2 \\
		\hline
		BM3 &4$	\times10^{-3}$     &1.6$	\times10^{-2}$ & 2.3$	\times10^{-1}$ & 3.8 \\
		\hline
	\end{tabular}

	\begin{tabular}{|c|c|c|c|}
	\hline
		& BR($\mu\rightarrow e \gamma$) & S & T    \\
		\hline
		\hline
		BM1 & 4.9$	\times10^{-23}$ & 4.3$	\times10^{-3}$& 3.0$	\times10^{-3}$   \\
		\hline
		BM2 &6.8$	\times10^{-28}$ & 2.7$	\times10^{-4}$& 3.2$	\times10^{-3}$   \\
		\hline
		BM3 &3.8$	\times10^{-28}$ & 6.9$	\times10^{-4}$& 3.2$	\times10^{-3}$   \\
		\hline
	\end{tabular}
	\caption{The top table shows that (BM1, BM2, BM3) satisfy current experimental
	constraints in the Higgs sector. The bottom one is used to illustrate how
	these solutions are in agreement with the bounds coming
	from LFV processes as well as the electroweak precision tests.}
	\label{tab:BM-II}
\end{table}

Finally, Table~\ref{tab:BM-III} shows the main DM annihilation channels in the model and prediction in the DM sector.

\begin{table}[t]
	\centering
	\begin{tabular}{|c|c|c|c|c|}
	\hline
		 &  BR($N_1 N_1 \to b \bar{b}$) & BR($ N_1 N_1 \to J J $) & BR($ N_1 N_1 \to h_2 h_2 $)  & BR($ N_1 N_1 \to Z Z $) \\
		\hline
		\hline
		BM1 & 7.8$	\times10^{-1}$ &9.8$	\times10^{-2}$ &- & -\\
		\hline
		BM2 & 5.9$	\times10^{-1}$ & 9.9$	\times10^{-2}$&- &2.34$	\times10^{-2}$\\
		\hline
		BM3 & 1.4$	\times10^{-5}$ & 1.4$	\times10^{-2}$&- &2.37$	\times10^{-1}$ \\
		\hline
	\end{tabular}

	\begin{tabular}{|c|c|c|c|c|}
	\hline
       & $BR(N_1 N_1 \to W^+ W^- $) & $\sigma_{\text{SI}}$[pb] & $\langle \sigma v \rangle_\gamma$ & $\Omega_c h^2$  	\\
		\hline
		\hline
		BM1 & -                   & 9.4$	\times10^{-12}$& 1.7$	\times10^{-34}$& 1.21$	\times10^{-1}$\\
		\hline
		BM2 &2$	\times10^{-1}$    & 3.4$	\times10^{-12}$& 1.0$	\times10^{-32}$& 1.20$	\times10^{-1}$	\\
		\hline
		BM3 & 4.7$	\times10^{-1}$& 2.2$	\times10^{-10}$& 4.5$	\times10^{-29}$& 1.19$	\times10^{-1}$ \\
		\hline
	\end{tabular}
	\caption{Main DM annihilation channels in the model. }
	\label{tab:BM-III}
\end{table}

\def\apj{Astrophys.~J.}                       
\def\apjl{Astrophys.~J.~Lett.}                
\def\apjs{Astrophys.~J.~Suppl.~Ser.}          
\def\aap{Astron.~\&~Astrophys.}               
\def\aj{Astron.~J.}                           %
\def\araa{Ann.~Rev.~Astron.~Astrophys.}       %
\def\mnras{Mon.~Not.~R.~Astron.~Soc.}         %
\def\physrep{Phys.~Rept.}                     %
\def\jcap{J.~Cosmology~Astropart.~Phys.}      
\def\jhep{J.~High~Ener.~Phys.}                
\def\prl{Phys.~Rev.~Lett.}                    
\def\prd{Phys.~Rev.~D}                        
\def\nphysa{Nucl.~Phys.~A}                    

\bibliography{refs}

\end{document}